\documentclass[aps,pra,reprint,superscriptaddress]{revtex4-2}

\usepackage{amsmath}
\usepackage{amssymb}
\usepackage{graphicx}
\usepackage{bm}
\usepackage{xcolor}
\usepackage[colorlinks=true,linkcolor=blue,citecolor=blue,urlcolor=blue]{hyperref}
\usepackage{float}
\usepackage{tcolorbox}
\usepackage{tikz}
\usepackage{ulem}
\usetikzlibrary{quantikz2}

\graphicspath{Figure/}

\newcommand{\ketbra}[2]{\left|#1\right\rangle\!\left\langle#2\right|}
\newcommand{\Tr}{\operatorname{Tr}}

\newcommand{\Rdel}{R_{\delta}}

\newcommand{\Anc}{\ensuremath{\mathcal{A}}}
\newcommand{\cD}{\ensuremath{\mathcal{D}}}
\newcommand{\cN}{\ensuremath{\mathcal{N}}}
\newcommand{\Sys}{\ensuremath{\mathcal{S}}}   
\newcommand{\Env}{\ensuremath{\mathcal{E}}}   
\newcommand{\Frac}{\ensuremath{\mathcal{F}}}   
\newcommand{\cI}{\ensuremath{\mathcal{I}}}

\newcommand{\hbin}{\mathrm{h}_2}

\begin{document}

\title{Quantum Darwinism through apparatus-mediated environmental recording}

\author{Don P.H Tran}
\affiliation{Faculty of Physics and Engineering Physics, University of Science, Ho Chi Minh City, 70000, Viet Nam}
\affiliation{Vietnam National University, Ho Chi Minh City, 70000, Viet Nam}

\author{Lan Nguyen Tran} 
\affiliation{Faculty of Physics and Engineering Physics, University of Science, Ho Chi Minh City, 70000, Viet Nam}
\affiliation{Vietnam National University, Ho Chi Minh City, 70000, Viet Nam}

\author{Le Bin Ho} 
\thanks{Electronic address: ho.bin.le.e3@tohoku.ac.jp}
\affiliation{Frontier Research Institute 
for Interdisciplinary Sciences, 
Tohoku University, Sendai 980-8578, Japan}
\affiliation{Department of Applied Physics, 
Graduate School of Engineering, 
Tohoku University, 
Sendai 980-8579, Japan}

\date{\today}

\begin{abstract}
Quantum Darwinism explains classical objectivity through redundant
environmental records of a system's pointer states.
We show that these records can be established through an intermediate
measurement apparatus without direct system--environment interactions.
Using exact analytical solutions of quantum circuit models,
we compare system monitoring with apparatus monitoring, where
the environment interacts only with the apparatus.
During encoding, system--apparatus correlations decrease as
system--environment correlations grow.
Without noise, both models yield identical accessible pointer
information and redundancy, while the discord accounts
for their different mutual information.
The reliability of these records depends on where noise acts.
Apparatus bit flips introduce errors shared by successive records,
limiting the information recoverable even from the complete
environment. In contrast, larger fragments can mitigate independent
noise on environment qubits after encoding.
Dephasing in the pointer basis leaves the record information
unchanged in the model considered.
These results extend the description of quantum Darwinism to
indirect measurements and clarify the conditions for reliable
environmental records.
\end{abstract}

\maketitle

\section{Introduction}

The emergence of classical behavior from quantum systems remains
a central question in quantum physics
\cite{Zurek2003RMP,Schlosshauer_2004}.
A central difficulty is understanding why macroscopic observations
yield definite outcomes when quantum theory allows superpositions,
as illustrated by Schr\"odinger's thought
experiment~\cite{Schrodinger_1935}.
One approach to this problem is decoherence theory, which explains
how interactions with the environment suppress interference
between components of a superposition
\cite{Zeh_1970,Joos_1985,Zurek_2005,schlosshauer_2019}.
These interactions select stable pointer states
\cite{Zurek_1981,Zurek_1982}, but the suppression of interference
does not explain why different observers independently obtain
the same information about the system.

Quantum Darwinism (QD) addresses this question through the
redundant encoding of pointer information in the
environment~\cite{Zurek2009NatPhys,BlumeKohout2006PRA}.
These records allow different observers to access separate
environmental fragments and infer the same pointer value
without directly measuring the system.
The environment thus both suppresses coherence and makes
pointer information available to multiple observers.
Experimental studies have investigated this mechanism in
photonic systems~\cite{Ciampini2018PRA} and a nitrogen-vacancy
center coupled to nuclear spins~\cite{Unden2019PRL}.
Recently, superconducting-circuit experiments have examined
the formation and retrieval of redundant environmental
records~\cite{Zhu2025SciAdv}.
Together, these studies test the environment's role in storing
and distributing pointer information.

Correlations between a system and an environmental fragment contain both classical information and quantum discord.
Quantum mutual information measures their total and does not
directly quantify the pointer information an observer can
read~\cite{zwolak_2013,Le2019PRL}.
Strong QD requires this mutual information to be entirely
accessible as classical information and to equal the system
entropy~\cite{Le2019PRL}.
While strong QD specifies the accessible information,
spectrum broadcast structure (SBS) describes records in which
different pointer values correspond to distinguishable
environmental states.
Observers can read these records without disturbing the system
or the records in other fragments~\cite{Horodecki2015PRA}.
In both approaches, decoherence alone does not establish classical
objectivity; observers must also independently identify the same
pointer value from separate fragments.

The spread of pointer information has also been studied in
composite quantum systems.
Cakmak et al. examined quantum discord and objectivity in
interacting two-qubit systems~\cite{Cakmak2021Entropy}.
Ryan and Paternostro investigated mediated redundancy, where
correlations allow information about one subsystem to reach
the environment through another~\cite{Ryan2022Entropy}.
The resulting records need not be ideal.
Studies of imperfect copying~\cite{Touil2022PRL} and nonideal
environments~\cite{Zwolak2010PRA} have examined their effects
on accessible information and redundant recording.
Together, these studies motivate examining the transfer of
pointer information through an intermediate apparatus and
the effects of noise on the resulting environmental records.

In many measurement settings, information reaches observers
through an intermediate apparatus.
For example, microwave photons trapped in a superconducting cavity
can be probed repeatedly by a stream of atoms without absorbing
the photons~\cite{Gleyzes_2007}.
Superconducting qubits are also read indirectly through a coupled
resonator whose resonance frequency depends on the qubit
state~\cite{blais_2004}.
Motivated by these settings, we consider a model in which
a system first becomes correlated with a measurement
apparatus.
The apparatus then interacts with the environment and transfers
system information to environmental fragments.
Figure~\ref{fig:architecture} shows this apparatus-monitoring
model alongside system monitoring.
We compare the two models using quantum mutual information,
Holevo information, and environmental redundancy.

We show that direct system--environment interactions are not required
to establish redundant pointer records.
As the environment acquires records through the apparatus,
the initial system--apparatus correlations are redistributed.
Without noise, apparatus monitoring preserves the accessible
pointer information and redundancy of system monitoring,
despite their different mutual information.
Noise exposes a limitation of this equivalence.
Apparatus bit flips introduce errors shared by successive records,
so collecting more records cannot remove the uncertainty
introduced during encoding.
In contrast, larger fragments can mitigate independent noise
on environment qubits after encoding.
Dephasing in the pointer basis leaves the record information
and redundancy unchanged at either location in the model considered.
These results extend quantum Darwinism to indirect measurements
and clarify the conditions for reliable environmental records.

The paper is organized as follows.
Section~\ref{sec:model} introduces the system-monitoring and
apparatus-monitoring models and their qubit circuits.
We present the analytical results in Sec.~\ref{sec:analytic}
and examine the effects of noise in Sec.~\ref{sec:noise},
followed by our conclusions in Sec.~\ref{sec:conclu}.

\section{Monitoring architectures and qubit model}
\label{sec:model}

\subsection{System monitoring}

\begin{figure}[t]
\centering
\includegraphics[width=\linewidth]{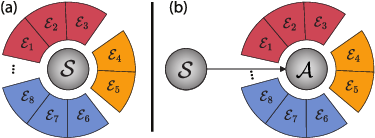}
\caption{
Two schemes for encoding system information into the environment.
(a) In system monitoring, separate environmental subsystems directly
record the pointer state of $\Sys$.
(b) In apparatus monitoring, $\Sys$ first transfers its pointer
information to $\Anc$.
This apparatus is then monitored by the environment without further
interaction with $\Sys$.
In both schemes, the environment
$\Env=\bigotimes_{k=1}^{N}\Env_k$ consists of $N$ subsystems,
and different observers access disjoint fragments.
}
\label{fig:architecture}
\end{figure}

We begin with the standard framework of
QD~\cite{Zurek2009NatPhys,BlumeKohout2006PRA}.
As illustrated in Fig.~\ref{fig:architecture}(a), a system
\Sys{} interacts directly with an environment
$\Env=\bigotimes_{k=1}^{N}\Env_k$.
The interaction correlates the system's pointer
states with conditional environmental records,
\begin{equation}
\ket{\Psi_{\Sys\Env}(t)}
=
\sum_s\sqrt{p_s}\ket{s}_{\Sys}\ket{\Env_s(t)},
\label{eq:pointer-branch-state}
\end{equation}
where $p_s=\Tr(\Pi_s\rho_{\Sys})$ and
$\Pi_{\Sys}=\{\Pi_0,\Pi_1\}$ denotes the pointer observable.
As the conditional records become distinguishable, coherence
between the pointer states is suppressed and information about
the pointer value becomes available in environmental fragments.

We divide the environment into an observed fragment $\Frac$
and its complement $\bar{\Frac}=\Env\setminus\Frac$.
The corresponding reduced states are
\begin{equation}
\rho_{\Sys}=\Tr_{\Env}(\rho_{\Sys\Env}),
\quad
\rho_{\Sys\Frac}=\Tr_{\bar{\Frac}}(\rho_{\Sys\Env}).
\label{eq:reduced-states}
\end{equation}
Different observers can access disjoint fragments without
interacting directly with \Sys{}.
The reduced state $\rho_{\Sys\Frac}$ determines both the
total system--fragment correlation and the pointer information
available from the fragment.

For a fragment $\Frac$, the pointer measurement defines the conditional
ensemble $\{p_s,\rho_{\Frac}^{s}\}$, where
$\rho_{\Frac}^{s}=\Tr_{\Sys}[(\Pi_s\otimes\openone)\rho_{\Sys\Frac}]/p_s$.
Here, $p_s$ is the probability of the pointer value $s$, and
$\rho_{\Frac}^{s}$ is the normalized conditional state of the fragment.
The quantum mutual information and Holevo quantity are
\begin{align}
\cI(\Sys:\Frac)
&=H(\rho_{\Sys})+H(\rho_{\Frac})-H(\rho_{\Sys\Frac}),
\label{eq:MI}\\
\chi(\Pi_{\Sys}:\Frac)
&=H(\rho_{\Frac})-\sum_s p_sH(\rho_{\Frac}^{s}),
\label{eq:holevo}
\end{align}
where $H(\rho)=-\Tr(\rho\log_2\rho)$ is the von Neumann entropy.
The mutual information $\cI(\Sys:\Frac)$ measures the total
correlations between $\Sys$ and $\Frac$, while the Holevo
quantity $\chi(\Pi_{\Sys}:\Frac)$ sets an upper limit on the
pointer information an observer can read from the fragment.
Their difference,
\begin{equation}
\cD(\Pi_{\Sys}:\Frac)
=
\cI(\Sys:\Frac)-\chi(\Pi_{\Sys}:\Frac),
\label{eq:discord}
\end{equation}
is the discord associated with the pointer
measurement~\cite{Ollivier2001PRL,Henderson2001JPA}.

The pointer entropy $H(\Pi_{\Sys})=-\sum_s p_s\log_2 p_s$
sets the maximum classical information available about the
pointer value.
For an allowed information loss $\delta$, we define
the minimum fragment size and redundancy as
\begin{align}
m_\delta
&=\min\left\{
m\in\{1,\ldots,N\}:
\overline{\chi}_m
\geq (1-\delta)H(\Pi_{\Sys})
\right\},\nonumber\\
\Rdel&=\frac{N}{m_\delta},
\label{eq:redundancy}
\end{align}
where $\overline{\chi}_m$ is the Holevo information averaged
over fragments of size $m$.
We assume $H(\Pi_{\Sys})>0$ and leave the redundancy undefined
if no fragment reaches the threshold.
A larger redundancy means that smaller fragments suffice
to reach the required fraction of pointer information,
according to the averaged Holevo criterion.

An ideal objective record has spectrum broadcast
structure~\cite{Horodecki2015PRA,Korbicz2021Quantum},
\begin{equation}
\rho_{\Sys\Env}^{\rm SBS}
=
\sum_i p_i\ket{s_i}\bra{s_i}\otimes
\left(\bigotimes_{k=1}^{N}\rho_{\Env_k}^{\,i}\right),
\label{eq:SBS}
\end{equation}
where records of different pointer values are distinguishable,
$\rho_{\Env_k}^{\,i}\rho_{\Env_k}^{\,j}=0$ for $i\neq j$.
Observers can read separate fragments and identify the same
pointer value without disturbing $\Sys$ or other observers'
records.

\subsection{Apparatus monitoring}

We consider an indirect monitoring architecture where the
environment interacts with an intermediate measurement apparatus
\Anc{} without directly interacting with \Sys{}, as illustrated
in Fig.~\ref{fig:architecture}(b).
The system first becomes correlated with \Anc{}, associating
each pointer state with an orthogonal apparatus state,
\begin{equation}
|s\rangle_{\Sys}|a_{\rm in}\rangle_{\Anc}
\longmapsto
|s\rangle_{\Sys}|a_s\rangle_{\Anc},
\quad
\langle a_s|a_{s'}\rangle=\delta_{ss'}.
\label{eq:general-SA-correlation}
\end{equation}
The environment then monitors \Anc{}, transferring pointer
information along the route
$\Sys\rightarrow\Anc\rightarrow\Env$.

This intermediate stage allows us to examine the redistribution
of the initial \Sys{}--\Anc{} correlations during environmental
recording and determine whether the resulting fragments provide
the same accessible pointer information as in system monitoring.

\subsection{Qubit model}
\label{sec:realization}

\begin{figure}[t]
\centering
\includegraphics[width=\linewidth]{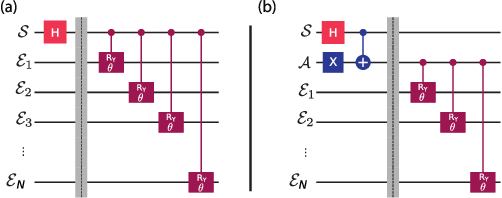}
\caption{
Qubit circuits for the two monitoring architectures.
(a) System monitoring: the system qubit sequentially controls
an $R_y(\theta)$ rotation on each environment qubit.
(b) Apparatus monitoring: the gates $H_{\Sys}$, $X_{\Anc}$, and
$\operatorname{CNOT}_{\Sys\rightarrow\Anc}$ correlate the system
with the apparatus.
The apparatus then controls the same rotations on the environment qubits.
}
\label{fig:circuit}
\end{figure}

We illustrate the two monitoring architectures with a qubit model,
where $\Sys$, $\Anc$, and each environmental subsystem $\Env_k$
are represented by qubits [Fig.~\ref{fig:circuit}].
The environment starts in $\ket{0}^{\otimes N}$,
and the model applies to any number $N$ of environment qubits.

In system monitoring [Fig.~\ref{fig:circuit}(a)], a Hadamard gate
prepares $\Sys$ in $(\ket{0}_{\Sys}+\ket{1}_{\Sys})/\sqrt{2}$.
It interacts with each $\Env_k$ qubit through
\begin{equation}
V_k^{(\Sys)}
=
\ketbra{0}{0}_{\Sys}\otimes\openone_{\Env_k}
+
\ketbra{1}{1}_{\Sys}\otimes R_y^{(\Env_k)}(\theta),
\label{eq:interaction}
\end{equation}
where $R_y(\theta)=\exp(-i\theta\sigma_y/2)$.
This interaction records the pointer information of $\Sys$
directly in each $\Env_k$ qubit.

In apparatus monitoring [Fig.~\ref{fig:circuit}(b)], \Env{}
does not interact with $\Sys$.
Starting from $\ket{0}_{\Sys}\ket{0}_{\Anc}$, the gates
$H_{\Sys}$, $X_{\Anc}$, and $\operatorname{CNOT}_{\Sys\rightarrow\Anc}$
prepare
\begin{equation}
\ket{\psi_{\Sys\Anc}}
=
\frac{\ket{0}_{\Sys}\ket{1}_{\Anc}
+\ket{1}_{\Sys}\ket{0}_{\Anc}}{\sqrt{2}}.
\label{eq:init}
\end{equation}
This state correlates each pointer state with an orthogonal
apparatus state.
The apparatus then controls the same rotation on each environment
qubit, transferring pointer information along
$\Sys\rightarrow\Anc\rightarrow\Env$.

We express the output states in terms of the environmental records
\begin{equation}
\ket{E_0}=\bigotimes_{k=1}^{N}\ket{0}_{\Env_k},
\quad
\ket{E_1}=\bigotimes_{k=1}^{N}R_y(\theta)\ket{0}_{\Env_k}.
\label{eq:environment-records}
\end{equation}
System monitoring gives
\begin{equation}
\ket{\Psi_{\Sys}}
=
\frac{\ket{0}_{\Sys}\ket{E_0}+\ket{1}_{\Sys}\ket{E_1}}{\sqrt{2}},
\label{eq:system-global}
\end{equation}
while apparatus monitoring gives
\begin{equation}
\ket{\Psi_{\Anc}}
=
\frac{\ket{0}_{\Sys}\ket{1}_{\Anc}\ket{E_1}
+\ket{1}_{\Sys}\ket{0}_{\Anc}\ket{E_0}}{\sqrt{2}}.
\label{eq:apparatus-global}
\end{equation}
The two architectures produce the same conditional environmental
records with exchanged pointer labels.
Their distinguishability and probabilities remain unchanged,
preserving the accessible pointer information.
This allows us to isolate the apparatus's role in redistributing
correlations.

For each $\Env_k$, the conditional records are $\ket{0}$ and
$\ket{\varepsilon}\equiv R_y(\theta)\ket{0}$, with overlap
\begin{equation}
\gamma=\langle0|\varepsilon\rangle=\cos\frac{\theta}{2}.
\label{eq:overlap}
\end{equation}
At $\theta=0$, the environment remains uncorrelated with the control,
while $\theta=\pi$ gives orthogonal records.
The interaction is non-demolition because it preserves
the control's pointer states.

\section{Results}
\label{sec:analytic}

We first compare the total correlations, accessible pointer
information, and redundancy in the two monitoring architectures.
We then examine the redistribution of the initial
\Sys{}--\Anc{} correlations as successive environment qubits
interact with \Anc{}.
Both architectures use the same environment size $N$ and
interaction angle $\theta$ to isolate the role of the
intermediate apparatus.
We evaluate redundancy using the minimum integer fragment size
defined in Eq.~\eqref{eq:redundancy}.

\subsection{Total and accessible information}
\label{sec:information}

Let $\Frac_m\subseteq\Env$ denote an observed fragment of $m$
environment qubits and $\overline{\Frac}_m=\Env\setminus\Frac_m$
its complement of $N-m$ qubits.
Each environment qubit undergoes the same controlled rotation
by $\theta$ [Fig.~\ref{fig:circuit}], so the conditional records
in the fragment and its complement have overlaps
\begin{equation}
\gamma_m=\cos^{m}\!\left(\frac{\theta}{2}\right),
\quad
\gamma_{N-m}=\cos^{N-m}\!\left(\frac{\theta}{2}\right).
\label{eq:fragment-overlaps}
\end{equation}
Their product gives the overlap of the full environmental records,
\begin{equation}
\gamma_N=\cos^{N}\!\left(\frac{\theta}{2}\right)
=\gamma_m\gamma_{N-m}.
\label{eq:total-environment-overlap}
\end{equation}

\begin{figure*}[t]
\centering
\includegraphics[width =0.8\linewidth]{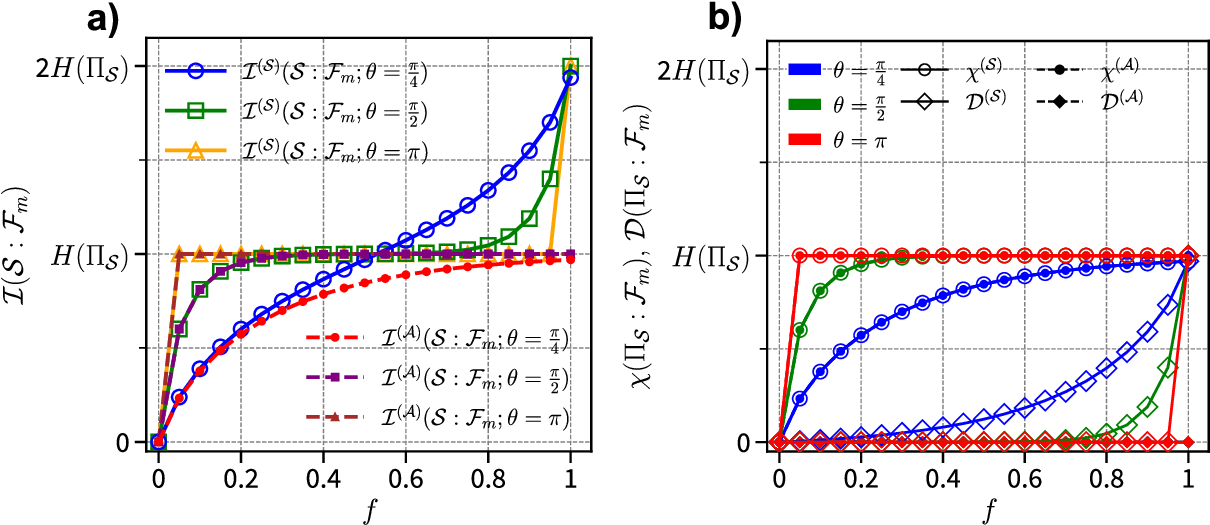}
\caption{
Comparison of system and apparatus monitoring for $N=20$.
(a) Mutual information $\cI(\Sys:\Frac_m)$ versus the fragment
fraction $f=m/N$.
(b) Holevo quantity $\chi(\Pi_{\Sys}:\Frac_m)$ (circles) and
pointer-basis discord $\cD(\Pi_{\Sys}:\Frac_m)$ (diamonds)
versus $f$.
Solid lines with open symbols denote system monitoring
($\Env$ interacts with $\Sys$), while dashed lines with filled
symbols denote apparatus monitoring ($\Env$ interacts with $\Anc$).
}
\label{fig:IchiD}
\end{figure*}

\subsubsection{Total correlations}

In system monitoring, the entropies in $\cI(\Sys:\Frac_m)$
depend on the record overlaps $\gamma_N$, $\gamma_m$, and
$\gamma_{N-m}$.
The global $\Sys\Env$ state is pure, so $\rho_{\Sys\Frac_m}$
and $\rho_{\overline{\Frac}_m}$ have the same nonzero eigenvalues
and equal entropies.
The overlap $\gamma_{N-m}$ of the complementary records thus
determines the entropy of $\rho_{\Sys\Frac_m}$.
Using the binary entropy
\begin{align}
\hbin(x)=-x\log_2x-(1-x)\log_2(1-x),
\end{align}
we obtain (Appendix~\ref{app:system-mi})
\begin{align}
\cI^{(\Sys)}(\Sys:\Frac_m)
&=\hbin\!\left(\frac{1+|\cos(\theta/2)|^{N}}{2}\right)
\nonumber\\
&\quad+\hbin\!\left(\frac{1+|\cos(\theta/2)|^{m}}{2}\right)
\nonumber\\
&\quad-\hbin\!\left(\frac{1+|\cos(\theta/2)|^{N-m}}{2}\right).
\label{eq:direct-mutual-information}
\end{align}

Equation~\eqref{eq:direct-mutual-information} gives zero for $m=0$,
since an empty fragment carries no information.
For the complete environment ($m=N$), the first two terms are equal
and the third vanishes because $\hbin(1)=0$, giving
\begin{equation}
\cI^{(\Sys)}(\Sys:\Env)
=
2\hbin\!\left(\frac{1+|\cos(\theta/2)|^{N}}{2}\right)
=
2H(\rho_{\Sys}).
\label{eq:complete-environment-mutual-information}
\end{equation}

The solid curves in Fig.~\ref{fig:IchiD}(a) show the system-monitoring
results.
The complementary-fragment term in
Eq.~\eqref{eq:direct-mutual-information} accounts for the final
rise of $\cI^{(\Sys)}(\Sys:\Frac_m)$ to $2H(\rho_{\Sys})$ at $m=N$.
For $\theta=\pi$, the complete environmental records are orthogonal,
so $\rho_{\Sys}$ is diagonal in the pointer basis with eigenvalues
$p_s=\Tr(\Pi_s\rho_{\Sys})$.
This gives $H(\rho_{\Sys})=H(\Pi_{\Sys})$, and the orange solid curve
reaches $2H(\Pi_{\Sys})$.
For the other interaction angles shown, the records remain
nonorthogonal and $\rho_{\Sys}$ retains coherence.
The corresponding curves end at
$2H(\rho_{\Sys})<2H(\Pi_{\Sys})$ even for the complete environment.
These results are consistent with previous
studies~\cite{Le2019PRL,zwolak_2013,Zurek2009NatPhys}.

In apparatus monitoring, tracing out the orthogonal apparatus states
removes coherence between the two system branches.
The reduced \Sys{}--$\Frac_m$ state is
\begin{equation}
\begin{aligned}
\rho_{\Sys\Frac_m}^{(\Anc)}
&=\frac12\ketbra{0}{0}_{\Sys}\otimes
\ketbra{e_{1,\Frac_m}}{e_{1,\Frac_m}}\\
&\quad+\frac12\ketbra{1}{1}_{\Sys}\otimes
\ketbra{e_{0,\Frac_m}}{e_{0,\Frac_m}}.
\end{aligned}
\label{eq:apparatus-fragment-cq}
\end{equation}
The conditional fragment states have overlap
$|\langle e_{0,\Frac_m}|e_{1,\Frac_m}\rangle|=|\cos(\theta/2)|^m$.
The state in Eq.~\eqref{eq:apparatus-fragment-cq} is
classical--quantum in the pointer basis.
Since its conditional fragment states are pure,
$H(\rho_{\Sys\Frac_m}^{(\Anc)})=H(\rho_{\Sys})$,
so the mutual information reduces to the fragment entropy,
\begin{equation}
\cI^{(\Anc)}(\Sys:\Frac_m)
=
\hbin\!\left(\frac{1+|\cos(\theta/2)|^m}{2}\right).
\label{eq:apparatus-mutual-information}
\end{equation}
Unlike Eq.~\eqref{eq:direct-mutual-information}, this expression
contains no contribution from the unobserved environment
$\overline{\Frac}_m$.
Even when the fragment includes the entire environment,
the apparatus remains unobserved.
For $m=N$, we obtain
\begin{equation}
\cI^{(\Anc)}(\Sys:\Env)
=
\hbin\!\left(\frac{1+|\cos(\theta/2)|^N}{2}\right),
\label{eq:apparatus-complete-environment-mutual-information}
\end{equation}
which reaches $H(\Pi_{\Sys})$ at $\theta=\pi$.

The dashed curves in Fig.~\ref{fig:IchiD}(a) show the
apparatus-monitoring results.
Unlike system monitoring, the mutual information depends
only on the distinguishability of the records in $\Frac_m$
and never exceeds $H(\Pi_{\Sys})$, even for the complete
environment.
Tracing out the unobserved apparatus removes the quantum
contribution to the \Sys{}--$\Frac_m$ mutual information.
The larger mutual information in system monitoring need
not imply more accessible pointer information.
We separate these contributions using the Holevo information
and discord in Fig.~\ref{fig:IchiD}(b).


\subsubsection{Accessible pointer information and discord}

To determine what an observer can learn about the pointer,
we decompose $\cI=\chi+\cD$.
Projecting the states in Eqs.~\eqref{eq:system-global} and
\eqref{eq:apparatus-global} onto the system pointer states
and tracing out the unobserved qubits gives the conditional
fragment states.

In system monitoring, pointer values $0$ and $1$ correspond
to $\ket{e_{0,\Frac_m}}=\ket{0_{\Frac_m}}$ and
$\ket{e_{1,\Frac_m}}$, respectively.
The corresponding ensemble is
\begin{equation}
\mathcal E_{\Frac_m}^{(\Sys)}
=
\left\{
\left(\frac12,\ketbra{e_{0,\Frac_m}}{e_{0,\Frac_m}}\right),
\left(\frac12,\ketbra{e_{1,\Frac_m}}{e_{1,\Frac_m}}\right)
\right\},
\label{eq:direct-fragment-ensemble}
\end{equation}
where each pair contains a pointer probability and the
fragment state conditioned on that value.
In apparatus monitoring, the same pointer values correspond
to the opposite records, giving
\begin{equation}
\mathcal E_{\Frac_m}^{(\Anc)}
=
\left\{
\left(\frac12,\ketbra{e_{1,\Frac_m}}{e_{1,\Frac_m}}\right),
\left(\frac12,\ketbra{e_{0,\Frac_m}}{e_{0,\Frac_m}}\right)
\right\}.
\label{eq:apparatus-fragment-ensemble}
\end{equation}

Both ensembles give the same average fragment state, with eigenvalues
$\lambda_m^{\pm}=(1\pm|\gamma_m|)/2$.
The conditional states are pure, so both Holevo quantities
equal the entropy of this average state,
\begin{align}
\chi^{(\Sys)}(\Pi_{\Sys}:\Frac_m)
&=\chi^{(\Anc)}(\Pi_{\Sys}:\Frac_m)
\nonumber\\
&=\hbin\!\left(\frac{1+|\cos(\theta/2)|^m}{2}\right).
\label{eq:qubit-holevo-equality}
\end{align}

This expression explains the coincident curves for $\chi$ in
Fig.~\ref{fig:IchiD}(b).
The interaction angle determines the distinguishability of the
fragment records, while switching the control from $\Sys$ to
$\Anc$ leaves their Holevo information unchanged.

The apparatus-monitoring state in
Eq.~\eqref{eq:apparatus-fragment-cq} is classical--quantum
in the pointer basis, giving
\begin{align}
\cI^{(\Anc)}(\Sys:\Frac_m)
&=\chi^{(\Anc)}(\Pi_{\Sys}:\Frac_m),
\nonumber\\
\cD^{(\Anc)}(\Pi_{\Sys}:\Frac_m)
&=0.
\label{eq:apparatus-information-identity}
\end{align}
For system monitoring, subtracting
Eq.~\eqref{eq:qubit-holevo-equality} from
Eq.~\eqref{eq:direct-mutual-information} gives
\begin{align}
\cD^{(\Sys)}(\Pi_{\Sys}:\Frac_m)
&=
\hbin\!\left(
\frac{1+|\cos(\theta/2)|^{N}}{2}
\right)
\nonumber\\
&\quad-
\hbin\!\left(
\frac{1+|\cos(\theta/2)|^{N-m}}{2}
\right).
\label{eq:direct-discord-closed}
\end{align}
This discord accounts for the difference between the two
mutual-information curves in Fig.~\ref{fig:IchiD}(a),
as shown in panel (b).
Both architectures retain the same Holevo information $\chi$ for
every fragment because their conditional records differ only
by an exchange of pointer labels.
The intermediate apparatus thus changes the quantum contribution
to the \Sys{}--$\Frac_m$ correlations while preserving the
classical pointer information encoded in the environment.

Vanishing discord does not guarantee that these fragments
contain pointer information.
At $\theta=0$, both $\cD$ and $\chi$
vanish, leaving no environmental record.

\subsubsection{Redundancy and objectivity}

Redundancy quantifies the number of disjoint environmental
fragments that independently provide nearly all the pointer
information, as defined in Eq.~\eqref{eq:redundancy}.
The two monitoring architectures have identical Holevo
information for every fragment size
[Eq.~\eqref{eq:qubit-holevo-equality}], so they reach the
information threshold at the same minimum fragment size.
This gives
\begin{equation}
m_\delta^{(\Sys)}=m_\delta^{(\Anc)},
\quad
R_\delta^{(\Sys)}=R_\delta^{(\Anc)}.
\label{eq:equal-redundancy}
\end{equation}
The intermediate apparatus changes the total \Sys{}--$\Frac_m$
correlations but preserves the redundancy of accessible pointer
information in the noiseless model.

\begin{figure*}[t]
\centering
\includegraphics[width= \linewidth]{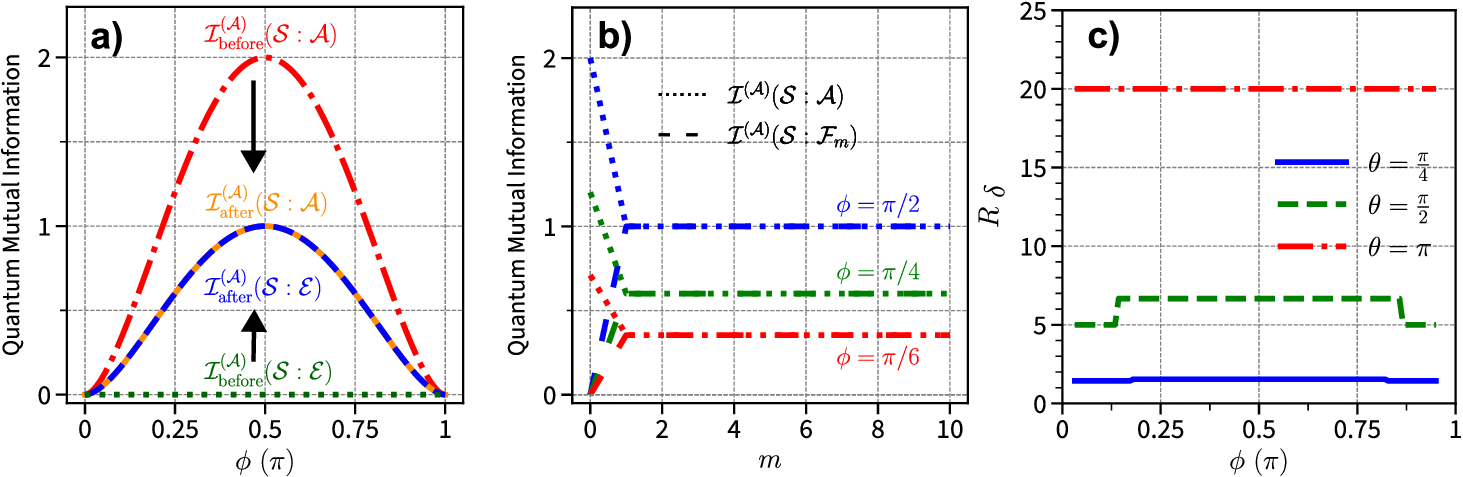}
\caption{
Information dynamics and redundancy in apparatus monitoring
with tunable \Sys{}--\Anc{} entanglement.
(a) Mutual information before and after environmental monitoring
versus the preparation angle $\phi$.
(b) Pairwise mutual information versus the number of environment qubits $m$ coupled to $\Anc$.
(c) Redundancy $R_\delta$ versus $\phi$ for $\delta=0.1$
and $\theta=\pi,\pi/2,\pi/4$.
}
\label{fig:Wigner_analysis}
\end{figure*}

\subsection{Information dynamics in apparatus monitoring}
\subsubsection{Correlation transfer to the environment}

We now examine the redistribution of the initial
\Sys{}--\Anc{} correlations as the apparatus transfers pointer
information to successive environment qubits.
To vary the initial pointer information, we replace the
Hadamard gate in Fig.~\ref{fig:circuit}(b) with
$R_y(\phi)_{\Sys}$, giving
\begin{equation}
R_y(\phi)\ket{0}_{\Sys}
=
\cos\!\left(\frac{\phi}{2}\right)\ket{0}_{\Sys}
+
\sin\!\left(\frac{\phi}{2}\right)\ket{1}_{\Sys}.
\label{eq:phi-preparation}
\end{equation}
After the \Sys{}--\Anc{} interaction, the joint state is
\begin{equation}
\ket{\psi_{\Sys\Anc}}
=
\cos\!\left(\frac{\phi}{2}\right)\ket{0}_{\Sys}\ket{1}_{\Anc}
+
\sin\!\left(\frac{\phi}{2}\right)\ket{1}_{\Sys}\ket{0}_{\Anc}.
\label{eq:phi-system-apparatus-state}
\end{equation}
The pointer probabilities are $p_0=\cos^2(\phi/2)$ and
$p_1=\sin^2(\phi/2)$, so the available pointer information is
\begin{equation}
H(\Pi_{\Sys})
=
-\sum_{s=0}^{1}p_s\log_2p_s
=
\hbin\!\left(\cos^2\frac{\phi}{2}\right).
\label{eq:Hpt}
\end{equation}

Figure~\ref{fig:Wigner_analysis}(a) shows the quantum mutual
information at $\theta=\pi$ as a function of $\phi$.
Here, ``before'' and ``after'' refer to the states before
and after all $N$ environment qubits have interacted with
\Anc{}, respectively.
The corresponding mutual information values are
\begin{align}
\cI_{\mathrm{before}}^{(\Anc)}(\Sys:\Anc)
&=2\hbin\!\left(\cos^2\frac{\phi}{2}\right)
=2H(\Pi_{\Sys}),\\
\cI_{\mathrm{after}}^{(\Anc)}(\Sys:\Anc)
&=\hbin\!\left(\cos^2\frac{\phi}{2}\right)
=H(\Pi_{\Sys}),\\
\cI_{\mathrm{before}}^{(\Anc)}(\Sys:\Env)
&=0,\\
\cI_{\mathrm{after}}^{(\Anc)}(\Sys:\Env)
&=\hbin\!\left(\cos^2\frac{\phi}{2}\right)
=H(\Pi_{\Sys}).
\label{eq:ledger}
\end{align}
Initially, the pure \Sys{}--\Anc{} state has mutual information
$2H(\Pi_{\Sys})$ (red curve), while the environment is uncorrelated
with \Sys{}, giving $\cI_{\mathrm{before}}^{(\Anc)}(\Sys:\Env)=0$.
Encoding at $\theta=\pi$ produces orthogonal environmental records,
reducing the \Sys{}--\Anc{} mutual information to $H(\Pi_{\Sys})$
(downward arrow) and increasing the \Sys{}--\Env{} mutual information
to the same value (upward arrow).
Both reduced states contain equal classical correlations,
giving the overlapping orange and blue curves.
At $\phi=\pi/2$, the initial \Sys{}--\Anc{} mutual information
is two bits, and both final values are one bit.

This redistribution satisfies
\begin{align}
&\cI_{\mathrm{before}}^{(\Anc)}(\Sys:\Anc)
+\cI_{\mathrm{before}}^{(\Anc)}(\Sys:\Env)
\nonumber\\
&\quad =
\cI_{\mathrm{after}}^{(\Anc)}(\Sys:\Anc)
+\cI_{\mathrm{after}}^{(\Anc)}(\Sys:\Env)
=2H(\Pi_{\Sys}).
\end{align}
During perfect encoding, the decrease in \Sys{}--\Anc{}
mutual information is exactly balanced by the increase
in \Sys{}--\Env{} mutual information.

Figure~\ref{fig:Wigner_analysis}(b) shows the redistribution
of these correlations as successive environment qubits interact
with \Anc{}, for perfect copying ($\theta=\pi$) and different
preparation angles $\phi$.
Initially ($m=0$), \Sys{} and \Anc{} share all correlations,
while the environment is uncorrelated with \Sys{}.
The first interaction creates orthogonal environmental records,
reducing $\cI^{(\Anc)}(\Sys:\Anc)$ from $2H(\Pi_{\Sys})$ to
$H(\Pi_{\Sys})$ and increasing $\cI^{(\Anc)}(\Sys:\Frac_m)$
from zero to the same value.
Both quantities remain unchanged for $m\geq1$ because subsequent
interactions create redundant records without adding pointer
information to the fragment.

More generally, this redistribution satisfies
\begin{equation}
\cI^{(\Anc)}(\Sys:\Anc)
+\cI^{(\Anc)}(\Sys:\Frac_m)
=2H(\Pi_{\Sys})
\end{equation}
at every interaction step in the noiseless model.
For imperfect copying ($0<\theta<\pi$), correlations transfer
gradually as more environment qubits interact with \Anc{}.
The interaction angle $\theta$ controls the transfer rate,
while the preparation angle $\phi$ sets the available
information through $H(\Pi_{\Sys})$.

\subsubsection{Redundancy of the environmental records}

We quantify redundancy using the Holevo information of a fragment
$\Frac_m$.
For an empty fragment, $\chi_0(\phi,\theta)=0$, while for $m\geq1$,
the conditional records have overlap
\begin{equation}
\left|\langle e_{0,\Frac_m}|e_{1,\Frac_m}\rangle\right|
=
\left|\cos\frac{\theta}{2}\right|^m.
\label{eq:fragment-overlap-phi-theta}
\end{equation}
The average fragment state has eigenvalues
\begin{align}
\lambda_m^{\pm}(\phi,\theta)
&=\frac{1}{2}\left[
1\pm\sqrt{1-\sin^2\phi
\left(1-\left|\cos\frac{\theta}{2}\right|^{2m}\right)}
\right].
\label{eq:fragment-eigenvalues-phi-theta}
\end{align}
Since the conditional states are pure, the Holevo information
equals the entropy of the average state,
\begin{equation}
\chi_m(\phi,\theta)
=
\hbin\!\left[\lambda_m^+(\phi,\theta)\right].
\label{eq:holevo-phi-theta}
\end{equation}

For an allowed information loss $\delta$, the target is
\begin{equation}
T_\delta(\phi)
=
(1-\delta)H(\Pi_{\Sys})
=
(1-\delta)\hbin\!\left(\cos^2\frac{\phi}{2}\right).
\label{eq:redundancy-target-phi}
\end{equation}
All fragments of size $m$ have the same Holevo information,
so $\overline{\chi}_m=\chi_m$.
The redundancy defined in Eq.~\eqref{eq:redundancy} then becomes
\begin{align}
m_\delta(\phi,\theta)
&=\min\left\{
m\in\{1,\ldots,N\}:
\chi_m(\phi,\theta)\geq T_\delta(\phi)
\right\},
\nonumber\\
R_\delta(\phi,\theta)
&=\frac{N}{m_\delta(\phi,\theta)}.
\label{eq:redundancy-phi-theta}
\end{align}
We assume $H(\Pi_{\Sys})>0$ and leave the redundancy undefined
if no fragment reaches the threshold.

Figure~\ref{fig:Wigner_analysis}(c) shows the redundancy
for $N=20$ and $\delta=0.1$.
As $\theta$ approaches $\pi$, the records become more
distinguishable, allowing smaller fragments to recover
$90\%$ of the pointer information and increasing redundancy.
At $\theta=\pi$, a single environment qubit carries the full
pointer information, giving $R_\delta=N=20$.
The preparation angle $\phi$ affects both the available
information and the fragment size needed to reach the threshold.
Since $m_\delta$ is an integer, the redundancy changes
in discrete steps.
In apparatus monitoring, $\cD=0$, so the mutual information
equals the Holevo information.

\section{Noise in environmental encoding}
\label{sec:noise}
\label{sec:beyond}

\begin{figure*}[t]
\centering
\includegraphics[width=\linewidth]{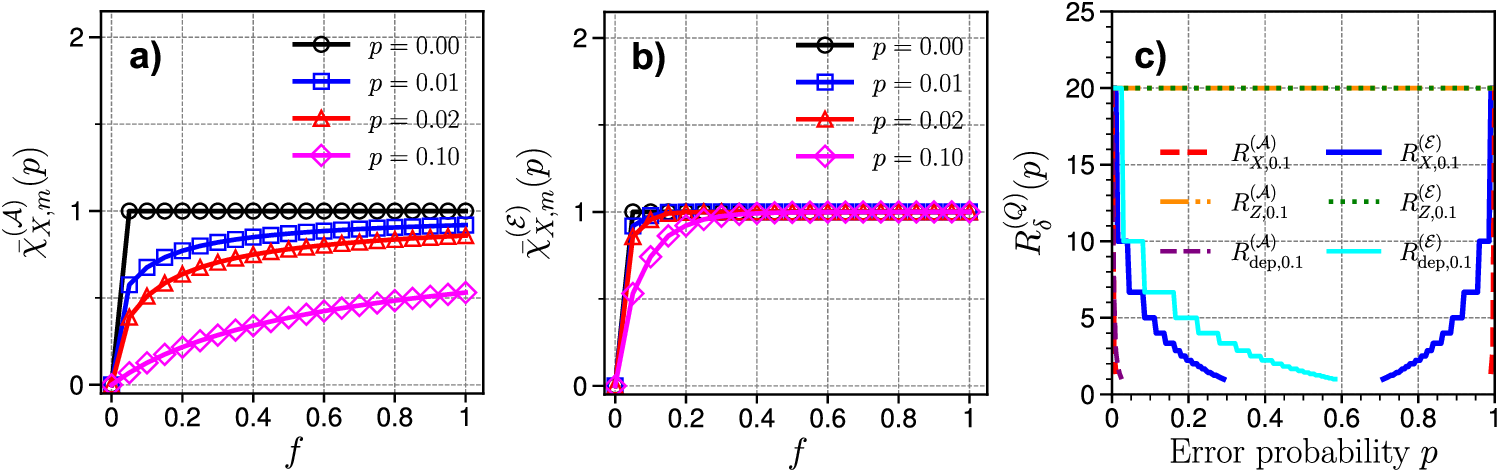}
\caption{
Effect of noise on environmental records for $N=20$,
$\theta=\pi$, and $\phi=\pi/2$.
(a) Average Holevo information versus fragment fraction $f=m/N$
for bit-flip noise on \Anc{} before each encoding interaction.
(b) The same quantity for bit-flip noise on each environment qubit
after encoding.
(c) Redundancy $R_\delta$ versus error probability $p$
for bit-flip ($X$), dephasing ($Z$), and depolarizing (dep)
noise at the two locations, with $\delta=0.1$.
}
\label{fig:Noise_comparison}
\end{figure*}

We examine the effects of noise on the information stored
in environmental records.
We set $N=20$, $\theta=\pi$, and $\phi=\pi/2$, so each noiseless
interaction creates orthogonal records of the pointer values.
We consider three channels acting on qubit $Q$,
\begin{align}
\cN_X^{(Q)}(\rho)
&=(1-p)\rho+p\sigma_x^{(Q)}\rho\sigma_x^{(Q)},
\nonumber\\
\cN_Z^{(Q)}(\rho)
&=(1-p)\rho+p\sigma_z^{(Q)}\rho\sigma_z^{(Q)},
\nonumber\\
\cN_{\mathrm{dep}}^{(Q)}(\rho)
&=(1-p)\rho
+p\left(\frac{\openone_Q}{2}\otimes\Tr_Q\rho\right),
\label{eq:noise-channels}
\end{align}
which describe bit-flip noise, dephasing in the pointer basis,
and depolarization, respectively.
Here, $p$ is the noise probability, and each channel acts
only on $Q$.
Apparatus noise acts on $Q=\Anc$ before each
\Anc{}--$\Env_k$ interaction.
Environmental noise acts once on $Q=\Env_k$ after that qubit
acquires its record.
We examine these two noise locations separately.

We start from
\begin{equation}
\rho^{(0)}
=
\ketbra{\psi_{\Sys\Anc}}{\psi_{\Sys\Anc}}
\otimes
\bigotimes_{k=1}^{N}\ketbra{0}{0}_{\Env_k},
\label{eq:noisy-global-initial-state}
\end{equation}
where $\ket{\psi_{\Sys\Anc}}$ is the \Sys{}--\Anc{} state
prepared in Eq.~\eqref{eq:init}.
Apparatus noise acts before each interaction, giving
\begin{equation}
\rho^{(k)}
=
U_{\Anc\Env_k}
\cN_{\mu}^{(\Anc)}\!\left(\rho^{(k-1)}\right)
U_{\Anc\Env_k}^{\dagger},
\label{eq:apparatus-noise-step}
\end{equation}
where $\mu\in\{X,Z,\mathrm{dep}\}$ and
$U_{\Anc\Env_k}
=\operatorname{CR}_y(\theta)_{\Anc\rightarrow\Env_k}$.
For bit-flip noise, the first step is
\begin{align}
\rho^{(1)}
={}&(1-p)U_{\Anc\Env_1}\rho^{(0)}
U_{\Anc\Env_1}^{\dagger}
\nonumber\\
&+p\,U_{\Anc\Env_1}\sigma_x^{(\Anc)}\rho^{(0)}
\sigma_x^{(\Anc)}U_{\Anc\Env_1}^{\dagger}.
\label{eq:first-apparatus-noise-step}
\end{align}
The first term describes no flip with probability $1-p$,
and the second describes a flip with probability $p$.

After all $N$ interactions, the conditional fragment states are
\begin{align}
\rho_{\Frac_m}^{s}
&=
\frac{
\Tr_{\Sys\Anc\overline{\Frac}_m}
\!\left[(\Pi_s\otimes\openone)\rho^{(N)}\right]
}{p_s},
\nonumber\\
p_s
&=
\Tr\!\left[(\Pi_s\otimes\openone)\rho^{(N)}\right],
\label{eq:noisy-conditional-fragment-state}
\end{align}
where $\overline{\Frac}_m=\Env\setminus\Frac_m$.
Since $\phi=\pi/2$ and no channel acts on \Sys{},
$p_0=p_1=1/2$.
The average fragment state is
\begin{equation}
\rho_{\Frac_m}
=
\frac{\rho_{\Frac_m}^{0}+\rho_{\Frac_m}^{1}}{2},
\label{eq:noisy-average-fragment-state}
\end{equation}
giving the Holevo information
\begin{align}
\chi(\Pi_{\Sys}:\Frac_m)
=H(\rho_{\Frac_m})
-\frac{H(\rho_{\Frac_m}^{0})+H(\rho_{\Frac_m}^{1})}{2}.
\label{eq:noisy-fragment-holevo-definition}
\end{align}

With apparatus noise, early and late records can carry different
amounts of pointer information.
We average the Holevo information over all fragments of size $m$,
\begin{equation}
\overline{\chi}_m
=
\frac{1}{\binom{N}{m}}
\sum_{\substack{\Frac_m\subseteq\Env\\|\Frac_m|=m}}
\chi(\Pi_{\Sys}:\Frac_m).
\label{eq:average-noisy-holevo}
\end{equation}
We denote this average by $\overline{\chi}_{\mu,m}^{(\Anc)}$
for apparatus noise and $\overline{\chi}_{\mu,m}^{(\Env)}$
for environmental noise.

For independent bit flips on \Anc{}, the probability that its
pointer value differs from the noiseless value after $k$ steps is
\begin{equation}
q_k=\frac{1-(1-2p)^k}{2}.
\label{eq:ancillary-error-probability}
\end{equation}
Let $i_1$ denote the earliest qubit index in $\Frac_m$.
Once this record is known, later records provide no additional
information about the original system pointer, giving
\begin{equation}
\chi(\Pi_{\Sys}:\Frac_m)=1-\hbin(q_{i_1}).
\label{eq:ancillary-noise-fragment-holevo}
\end{equation}
There are $\binom{N-k}{m-1}$ fragments whose earliest qubit
is $\Env_k$.
Averaging over these fragments gives
\begin{align}
\overline{\chi}^{(\Anc)}_{X,m}(p)
={}&\frac{1}{\binom{N}{m}}
\sum_{k=1}^{N-m+1}\binom{N-k}{m-1}
\nonumber\\
&\times
\left\{
1-\hbin\!\left[\frac{1-(1-2p)^k}{2}\right]
\right\}.
\label{eq:averaged-ancillary-noise-holevo}
\end{align}
For the complete environment, only $k=1$ contributes, so
\begin{equation}
\overline{\chi}^{(\Anc)}_{X,N}(p)=1-\hbin(p).
\label{eq:complete-environment-ancillary-noise}
\end{equation}

Figure~\ref{fig:Noise_comparison}(a) shows the averaged
Holevo curves, which recover the noiseless result at $p=0$.
As $p$ approaches $1/2$, the pointer information decreases
even when the full environment is observed.
For sufficiently strong noise, the curves remain below
the $90\%$ redundancy threshold, as also indicated by
the red dashed curve in panel (c).

For environmental noise, the apparatus first writes a record,
and the channel then acts on $\Env_k$,
\begin{equation}
\rho^{(k)}
=
\cN_{\mu}^{(\Env_k)}\!\left(
U_{\Anc\Env_k}\rho^{(k-1)}U_{\Anc\Env_k}^{\dagger}
\right).
\label{eq:storage-noise-step}
\end{equation}
The noise acts independently on each environment qubit.
For bit-flip noise, each bit string with $l$ ones in a fragment
of size $m$ has probability
\begin{equation}
P_l
=
\frac12\left[
(1-p)^l p^{m-l}
+
(1-p)^{m-l}p^l
\right].
\end{equation}
There are $\binom{m}{l}$ such strings, giving the Holevo information
\begin{equation}
\overline{\chi}^{(\Env)}_{X,m}(p)
=
-\sum_{l=0}^{m}\binom{m}{l}P_l\log_2P_l
-
m\hbin(p).
\end{equation}

Figure~\ref{fig:Noise_comparison}(b) shows that larger fragments
recover pointer information despite independent environmental noise.
At $p=0.10$, a fragment of $m=4$ qubits ($f=0.2$)
exceeds the $90\%$ threshold.
Environmental noise acts after encoding and leaves the apparatus
unchanged, so it does not affect subsequent records.
For fixed $p\neq1/2$, combining more records gives
\begin{equation}
\lim_{m\to\infty}\overline{\chi}^{(\Env)}_{X,m}(p)=1.
\end{equation}
At $p=1/2$, each record is independent of the pointer,
and the Holevo information vanishes.

We evaluate redundancy using the minimum integer fragment size
defined in Eq.~\eqref{eq:redundancy-phi-theta}.
For noise location $Q\in\{\Anc,\Env\}$ and channel $\mu$,
\begin{align}
m_{\mu,\delta}^{(Q)}(p)
&=
\min\left\{
m\in\{1,\ldots,N\}:
\overline{\chi}_{\mu,m}^{(Q)}(p)\geq1-\delta
\right\},
\nonumber\\
R_{\mu,\delta}^{(Q)}(p)
&=
\frac{N}{m_{\mu,\delta}^{(Q)}(p)}.
\label{eq:redundancy-bitflip}
\end{align}
Here, $\delta=0.1$ and $H(\Pi_{\Sys})=1$.
Dephasing in the pointer basis leaves the conditional record
populations unchanged, giving
\begin{equation}
\overline{\chi}_{Z,m}^{(Q)}(p)=1
\quad (m\geq1),
\quad
R_{Z,\delta}^{(Q)}(p)=N
\label{eq:redundancy-dephasing}
\end{equation}
at either noise location.
For these pointer records, depolarization acts as a bit flip
with probability $p/2$, so
\begin{align}
\overline{\chi}_{\mathrm{dep},m}^{(Q)}(p)
&=
\overline{\chi}_{X,m}^{(Q)}(p/2),
\nonumber\\
R_{\mathrm{dep},\delta}^{(Q)}(p)
&=
R_{X,\delta}^{(Q)}(p/2).
\label{eq:redundancy-depolarizing}
\end{align}

Figure~\ref{fig:Noise_comparison}(c) compares the resulting
redundancies.
Dephasing leaves them unchanged, while bit-flip and depolarizing
noise reduce them.
The bit-flip curves are symmetric about $p=1/2$ and recover
at $p=1$, where deterministic flips preserve the pointer information.
For depolarizing noise, redundancy does not increase with $p$.
The integer fragment size produces discrete steps in the curves.

Apparatus noise reduces redundancy more strongly than independent
environmental noise in the cases considered.
An apparatus flip changes the pointer value copied into successive
records, introducing shared errors.
An environmental error affects only one record, allowing observers
to recover pointer information by combining independent records.

\section{Conclusion}
\label{sec:conclu}

We have studied the transfer, accessibility, and robustness
of pointer information through an intermediate measurement
apparatus in quantum Darwinism.
As the environment acquires records, the initial
\Sys{}--\Anc{} correlations are redistributed.
Without noise, apparatus monitoring yields the same accessible
pointer information and redundancy as system monitoring,
despite their different mutual information.
Direct \Sys{}--\Env{} interactions are thus not required
to establish redundant pointer records.

The reliability of these records depends on where noise acts.
Apparatus bit flips introduce errors shared by successive
records, limiting the recoverable pointer information.
In contrast, larger fragments can mitigate independent noise
on environment qubits after encoding.
Dephasing in the pointer basis leaves the accessible record
information unchanged in the model considered.
These results extend the description of quantum Darwinism
to indirect measurements and clarify the conditions for
reliable environmental records.
Future work could examine these conditions under more realistic
open-system dynamics and imperfect measurement interactions.

\section*{Acknowledgments}
This work is supported by the Tohoku Initiative for Fostering Global Researchers for Interdisciplinary Sciences (TI-FRIS) of MEXT's Strategic Professional Development Program for Young Researchers. L. N. T is funded by Vietnam National University,  Ho Chi Minh City (VNU-HCM) under grant number B2026-18-18. 

\section*{Data availability} 
The data that support the findings of this article are openly available~\cite{git}.

\appendix

\section{Derivation of the system-monitoring mutual information}
\label{app:system-mi}

We derive Eqs.~\eqref{eq:direct-mutual-information} and
\eqref{eq:complete-environment-mutual-information}.
In system monitoring, the joint $\Sys\Env$ state is
\begin{equation}
\ket{\Psi_{\Sys\Env}}
=\frac{1}{\sqrt{2}}
\left(
\ket{0}_{\Sys}\ket{E_0}
+\ket{1}_{\Sys}\ket{E_1}
\right),
\label{eq:app-global-system-state}
\end{equation}
where the conditional environmental records are
\begin{equation}
\ket{E_0}=\bigotimes_{k=1}^{N}\ket{0}_{\Env_k},
\quad
\ket{E_1}=\bigotimes_{k=1}^{N}R_y(\theta)\ket{0}_{\Env_k},
\label{eq:environment-records_app}
\end{equation}
as defined in Eq.~\eqref{eq:environment-records}.

We divide the environment into an observed fragment $\Frac_m$
of $m$ qubits and its complement $\overline{\Frac}_m$.
The records factorize as
\begin{equation}
\ket{E_s}
=\ket{e_{s,\Frac_m}}\ket{e_{s,\overline{\Frac}_m}},
\quad s=0,1.
\label{eq:app-record-factorization}
\end{equation}
Each environment qubit undergoes the same controlled rotation,
giving the overlaps
\begin{align}
\langle e_{0,\Frac_m}|e_{1,\Frac_m}\rangle
&=\gamma_m=\cos^m\!\left(\frac{\theta}{2}\right),
\nonumber\\
\langle e_{0,\overline{\Frac}_m}|e_{1,\overline{\Frac}_m}\rangle
&=\gamma_{N-m}=\cos^{N-m}\!\left(\frac{\theta}{2}\right).
\label{eq:app-fragment-overlaps}
\end{align}
Their product gives the overlap of the complete records,
\begin{equation}
\langle E_0|E_1\rangle
=\gamma_N=\gamma_m\gamma_{N-m}
=\cos^N\!\left(\frac{\theta}{2}\right).
\label{eq:app-total-overlap}
\end{equation}

To calculate the entropies, we first consider an equally weighted
mixture of two normalized states,
\begin{equation}
\rho_{uv}=\frac{1}{2}
\Big(\ketbra{u}{u}+\ketbra{v}{v}\Big).
\label{eq:app-two-state-mixture}
\end{equation}
Its spectrum is determined by
\begin{equation}
\lambda_{\pm}=\frac{1\pm|\langle u|v\rangle|}{2},
\label{eq:app-two-state-spectrum}
\end{equation}
with any remaining eigenvalues equal to zero.
These values follow from diagonalizing the Gram matrix
\begin{equation}
G=\frac{1}{2}
\begin{pmatrix}
1 & \langle u|v\rangle\\
\langle v|u\rangle & 1
\end{pmatrix},
\label{eq:app-gram-matrix}
\end{equation}
which has the same nonzero eigenvalues as $\rho_{uv}$.
The entropy is
\begin{equation}
H(\rho_{uv})
=\hbin\!\left(\frac{1+|\langle u|v\rangle|}{2}\right).
\label{eq:app-two-state-entropy}
\end{equation}

Tracing Eq.~\eqref{eq:app-global-system-state} over the environment
gives the reduced system state in the pointer basis,
\begin{equation}
\rho_{\Sys}=\frac{1}{2}
\begin{pmatrix}
1 & \gamma_N^{*}\\
\gamma_N & 1
\end{pmatrix}.
\label{eq:app-reduced-system}
\end{equation}
Its eigenvalues are $(1\pm|\gamma_N|)/2$, giving
\begin{equation}
H(\rho_{\Sys})
=\hbin\!\left(\frac{1+|\gamma_N|}{2}\right).
\label{eq:app-system-entropy}
\end{equation}
Tracing over $\Sys\overline{\Frac}_m$ gives
\begin{equation}
\rho_{\Frac_m}=\frac{1}{2}
\Big(
\ketbra{e_{0,\Frac_m}}{e_{0,\Frac_m}}
+\ketbra{e_{1,\Frac_m}}{e_{1,\Frac_m}}
\Big).
\label{eq:app-reduced-fragment}
\end{equation}
Using its record overlap in Eq.~\eqref{eq:app-two-state-entropy},
we obtain
\begin{equation}
H(\rho_{\Frac_m})
=\hbin\!\left(\frac{1+|\gamma_m|}{2}\right).
\label{eq:app-fragment-entropy}
\end{equation}

The global state is pure, so the reduced states on
$\Sys\Frac_m$ and $\overline{\Frac}_m$ have the same nonzero
eigenvalues and equal entropies,
\begin{equation}
H(\rho_{\Sys\Frac_m})=H(\rho_{\overline{\Frac}_m}).
\label{eq:app-complementary-entropies}
\end{equation}
The unobserved environment has the state
\begin{equation}
\rho_{\overline{\Frac}_m}=\frac{1}{2}
\left(
\ketbra{e_{0,\overline{\Frac}_m}}{e_{0,\overline{\Frac}_m}}
+\ketbra{e_{1,\overline{\Frac}_m}}{e_{1,\overline{\Frac}_m}}
\right).
\label{eq:app-reduced-complement}
\end{equation}
Its record overlap $\gamma_{N-m}$ determines the joint entropy,
\begin{equation}
H(\rho_{\Sys\Frac_m})
=\hbin\!\left(\frac{1+|\gamma_{N-m}|}{2}\right).
\label{eq:app-joint-entropy}
\end{equation}

Substituting Eqs.~\eqref{eq:app-system-entropy},
\eqref{eq:app-fragment-entropy}, and \eqref{eq:app-joint-entropy}
into
\begin{equation}
\cI^{(\Sys)}(\Sys:\Frac_m)
=H(\rho_{\Sys})+H(\rho_{\Frac_m})-H(\rho_{\Sys\Frac_m})
\label{eq:app-mutual-information-definition}
\end{equation}
gives
\begin{align}
\cI^{(\Sys)}(\Sys:\Frac_m)
&=\hbin\!\left(\frac{1+|\gamma_N|}{2}\right)
+\hbin\!\left(\frac{1+|\gamma_m|}{2}\right)
\nonumber\\
&\quad-\hbin\!\left(\frac{1+|\gamma_{N-m}|}{2}\right).
\label{eq:app-mutual-information-overlaps}
\end{align}
The overlaps in Eqs.~\eqref{eq:app-fragment-overlaps} and
\eqref{eq:app-total-overlap} recover
Eq.~\eqref{eq:direct-mutual-information}.

For an empty fragment ($m=0$), $\gamma_m=1$ and
$\gamma_{N-m}=\gamma_N$.
Since $\hbin(1)=0$, the first and third terms cancel, giving
\begin{equation}
\cI^{(\Sys)}(\Sys:\Frac_0)=0.
\label{eq:app-empty-fragment}
\end{equation}
For the complete environment ($m=N$), $\gamma_m=\gamma_N$
and $\gamma_{N-m}=1$, so
\begin{equation}
\cI^{(\Sys)}(\Sys:\Env)
=2\hbin\!\left(\frac{1+|\gamma_N|}{2}\right)
=2H(\rho_{\Sys}).
\label{eq:app-complete-environment}
\end{equation}
This recovers the identity $\cI(\Sys:\Env)=2H(\rho_{\Sys})$
for a pure bipartite state.

\bibliography{refs}

\end{document}